# Explainable AI: Definition and attributes of a good explanation for health AI

Evangelia Kyrimi[1,*], Scott McLachlan[2], Jared M Wohlgemut[3,4], Zane B Perkins[3,4], David A. Lagnado[5], William Marsh[1] and the ExAIDSS Expert Group (mention at the end)

[1] School of Electronic Engineering and Computer Science, Queen Mary University of London, London, UK
[2] School of Nursing, Midwifery and Palliative Care, King's College London, London, UK
[3] Centre for Trauma Sciences, Blizard Institute, Queen Mary University of London, London, UK
[4] Royal London Hospital, Barts Health NHS Trust, London, UK
[5] Department of Experiment Psychology, University College London, London, UK

[*] *Corresponding author e-mail address:* <u>e.kyrimi@qmul.ac.uk</u> *(Twitter handle: @LinaKyrimi)*

**Abstract**
Proposals of artificial intelligence (AI) solutions based on increasingly complex and accurate predictive models are becoming ubiquitous across many disciplines. As the complexity of these models grows, transparency and users' understanding often diminish. This suggests that accurate prediction alone is insufficient for making an AI-based solution truly useful. In the development of healthcare systems, this introduces new issues related to accountability and safety. Understanding how and why an AI system makes a recommendation may require complex explanations of its inner workings and reasoning processes. Although research on explainable AI (XAI) has significantly increased in recent years and there is high demand for XAI in medicine, defining what constitutes a good explanation remains ad hoc, and providing adequate explanations continues to be challenging. To fully realize the potential of AI, it is critical to address two fundamental questions about explanations for safety-critical AI applications, such as health-AI: (1) What is an explanation in health-AI? and (2) What are the attributes of a good explanation in health-AI? In this study, we examined published literature and gathered expert opinions through a two-round Delphi study. The research outputs include (1) a definition of what constitutes an explanation in health-AI and (2) a comprehensive list of attributes that characterize a good explanation in health-AI.

# 1. Introduction

Artificial Intelligence (AI) has potential for transformative impact in a number of fields and industries. The International Data Corporation (IDC) predicts annual AI spending will exceed $300bn by 2026 as more companies integrate these intelligent technologies into their product and service offerings [1]. Some of the key applications include:

(i) Healthcare: for disease diagnosis and personalized treatment recommendation [2].
(ii) Finance: for algorithmic trading, fraud detection and predicting investment decisions [3].

    (iii) Autonomous Vehicles: for advanced driver assistance systems and self-driving car technologies [4].
    (iv) Natural Language Processing (NLP): for language translation, speech recognition and voice assistants like Siri and Alexa [5].
    (v) Cybersecurity: for analysing patterns, identifying anomalies and automating security processes [6].
    (vi) Entertainment: for music and video generation, virtual reality simulations and game development [7], [8].
    (vii) Legal Services: for legal research, contract analysis and document review [8].

Despite all the successes of AI, recent work shows that AI can unintentionally harm humans and that it is precisely the large-scale and wide introduction of AI technologies that hold enormous and unimagined potential for new types of unforeseen threats [9]. The absence of an understanding of how an AI works can allow both unintentional and deliberate bias to shape the recommendations, predictions and decisions AI may be used to make [9]. For example, AI can make unreliable decisions in safety-critical scenarios (e.g. in the medical domain) or undermine fairness by inadvertently discriminating against a group [10]. For this reason, explainable AI (XAI) is essential to ensure transparency, fairness, and ethical integrity in AI-driven decision-making in order to minimise the potential harms of unchecked AI implementation [11], [12], [13]. By providing insights into AI systems' inner workings, explainability empowers trust, understanding, and effective leveraging of AI technologies for positive societal impact [14], [15]. An increasing requirement for an explanation has arisen due to the broad adoption of machine learning (ML) approaches, where the reasoning task is often performed in what are known as *blackbox systems* where it is unclear why a specific result has been reached [16], [17]. Increasing use of these systems in healthcare, where important decisions about humans are made, raises new issues for accountability and safety [14]. Explainable and trustworthy AI have also been included in the European Commission's ethics guidelines [18].

Without a solution to the problem of trustworthy AI and user acceptance of healthcare technologies generally, the benefits of these systems will never be realised and all our efforts to develop accurate health-AI will be in vain [19]. This research aimed to answer two fundamental questions of explanation in health-AI that remain unanswered:
1) What is an explanation in health-AI?
2) What are the attributes of a good explanation in health-AI?

To achieve the above objectives, we consulted experts from diverse backgrounds using a Delphi study. The research outputs are (1) a definition and (2) a global list of attributes of a good explanation in health-AI. By understanding the attributes of a good explanation, improved explanation algorithms will be developed. Generating explanations is not enough, it is also crucial to evaluate how good these explanations are. Thus, the proposed list of attributes will serve as a means to produce an evaluation process, which is lacking with respect to formalised measures.

The remainder of this paper is organized as follows: Background information is presented in section 2. The methodology and results are explained in Sections 3 and 4, respectively. Finally, a detailed discussion and a study conclusion are presented in Sections 5 and 6.

# 2. Background and Motivation

## 2.1 Law and Ethics

Medical AI is considered a high-risk AI application in the proposed European legislation [14]. Therefore, having an explanation for why and how some conclusions were reached by the system is extremely important [20]. In healthcare, XAI is urgently needed for many purposes including education, research and clinical decision-making [15], [21]. If medical professionals are complemented by sophisticated AI systems, and in some cases even overruled, the human experts must, on demand, still have a chance to understand and to retrace the machine decision process [14], [22]. A key requirement for adopting these systems is that users must feel confident in their recommendations. The reasons for equipping intelligent systems with explanatory capabilities are not limited to issues of user rights and of technology acceptance, though. Explainability is also required by designers and developers to enhance system robustness and to enable diagnostics to prevent bias, unfairness, and discrimination, as well as to increase trust by all users in why and how decisions are made [23]. Given that this issue is widely acknowledged, and regulations have been put in place to restrict the use of ambiguous health AI [24], continued research towards a reliable approach for explanation development is essential.

When the European Union (EU) adopted the General Data Protection Regulation (GDPR) in 2016, they granted European citizens a *right to explanation* if they are affected by algorithmic decisions [25]. Research on XAI has significantly increased since GDPR came into force in the EU in 2018 [25], and the demand for XAI in medicine is high [26]. However, determining what constitutes a good explanation is *ad hoc* and providing adequate explanations remains a challenge [15], [23]. For example, if an AI system rejects an individual's application for a loan, the applicant is entitled to request the justifications that led to the decision so that they may ensure consistency with other laws, regulations and rights. In early 2017, the Defense Advanced Research Projects Agency (DARPA) funded the *Explainable AI (XAI) Program*, which resulted in the term XAI gaining popularity in the research community [27], even though the first use of the abbreviation XAI for the term *explainable artificial intelligence* dated to 2004 [28]. In 2019, the EU published the *Ethics Guidelines for Trustworthy AI* which includes a general framework where explainability is an integral component [29]. These guidelines formed the basis for several sections of the *Artificial Intelligence Act* proposed by the European Commission (EC) in April 2021. Finally, the 2023 UK white paper '*A pro-innovation approach to AI regulation*' identifies *appropriate transparency and explainability* as one of the key principles for developing responsible AI. The need for transparent AI led to a significant increase in the size of the XAI community over the last few years [20], [30].

## 2.2 Related work on Explainable AI (XAI)

There is an extensive body of literature reviewing different aspects of XAI. Much of this literature focuses heavily on how we can classify explanation methods [31], [32], [33], [34] as well as how XAI can impact affected parties [35], [36], [37]. Others also review evaluation methodologies for explainable systems [24]. Papers on foundational aspects of an explanation are scarce. Defining what an explanation is and what constitutes a good explanation remains *ad hoc* [15], [23].

The majority of the published papers that focus on XAI definition either review existing literature [30], [38], [39] or seek definitions inspired by existing literature and their own expertise [16], [17], [33], [40], [41], [32], [42], [43], [44], [45], [46]. Existing definitions of XAI often fail to provide specific guidance. These definitions, which may be at least in part correct, tell us very little about what an explanation is. They omit aspects such as the semantic entity of an explanation, the aim of the explaining inference, the target audience, etc. In addition, an increasing number of contributions tend to rely on their own, often intuitive, notions of "explainability". This can lead to a failure to provide satisfactory explanations. Finally, defining what explainability in AI is should be an interdisciplinary task, yet the existing work focuses on a single discipline.

Determining the criteria for a good explanation is an area of interest in various scientific fields [47]. Many researchers propose a list of attributes for a good explanation based merely on literature synthesis [20], [23], [39], [48], [49], [50], [51], [52] [53], [54], [55], [56]. Others focus only on their own experience and expertise [46], [57]. There are few papers that draw on the results of user experiments in which they investigate how AI users perceive a proposed explanation and, based on their responses, infer explanation characteristics [35], [58], [59], [60], [61]. Affected parties are rarely involved directly in the exploration of the characteristics needed for a good explanation. Discussions about explainability should involve many different groups including the AI community, AI users, human-computer interaction researchers, researchers from the social sciences, and regulators. However, existing works are based on a restricted point of view, missing the multidisciplinary aspect of the problem.

## 3. Methodology

To achieve this study's objectives, published literature and expert opinions, using a Delphi study of participants with a diverse research background, were analysed (Figure 1).

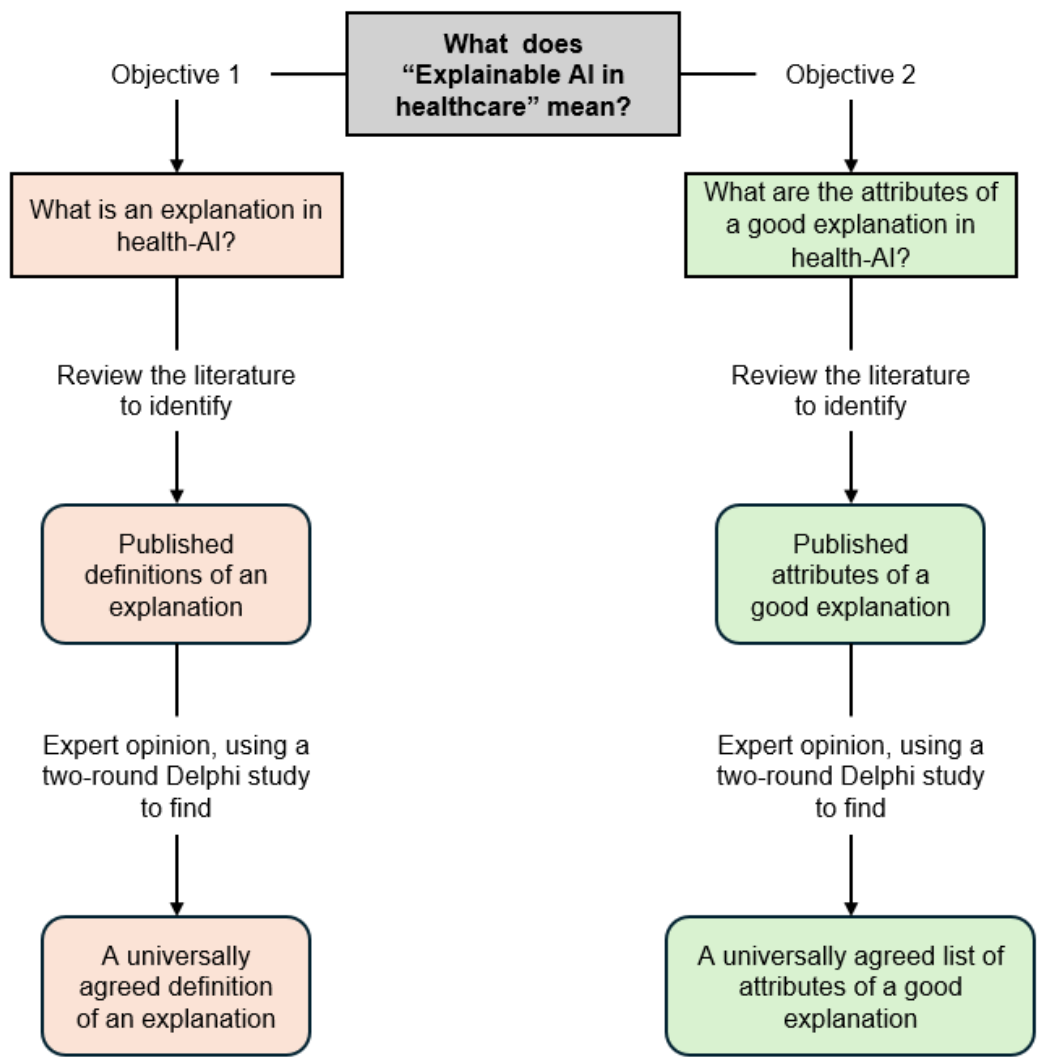

**Figure 1.** Framework for identifying what it means for AI to be explainable in healthcare.

### 3.1 Scoping review

A search of major databases including PubMed, MedLine, ScienceDirect, Scopus, DOAJ and Elsevier was performed using the following search query:

*[(XAI OR "explainable AI") AND (definition OR criteria OR desiderata OR attribute OR characteristic OR feature OR notion)]*

Due to the high number of returned articles, further scrutiny was necessary to identify the most relevant articles for this study. We selected papers for inclusion where the described keywords were present in the title or abstract. Additional screening was performed to exclude papers: (i) published before 2018 (when GDPR came into effect in the EU); (ii) not published in English; (iii) where full access to the paper was not possible; or (iv) articles that had not been through peer review. The remaining papers were those meeting the inclusion (IC) without triggering the exclusion (EC) criteria. These criteria are presented in Table 1. The results from the scoping review were used to form the questionnaire for the first round of Delphi study.

**Table 1.** Inclusion and exclusion criteria

| | |
|---|---|
| **Inclusion Criteria** | |
| IC-1: | Review papers that synthesise existing definitions of explanations in AI and/or summarise previously published attributes for a good explanation. |
| IC-2: | Papers that propose a new definition of explanation in AI and/or a list of attributes for a good explanation. |
| **Exclusion Criteria** | |
| EC-1: | Papers that define what an explanation is from a philosophical point of view without considering explanations in AI. |
| EC-2: | Papers that explore XAI aspects such as algorithms, evaluation metrics etc without exploring the two fundamental questions studied in this review. |
| EC-3: | Papers that only mention definitions of explanation or attributes of a good explanation in AI that are published elsewhere. |
| EC-4: | There is a statement in the title/abstract that a definition or attributes of an explanation in AI is provided, but none is detailed. |
| EC-5: | Papers that study how people explain/ justify their decisions. |

We also used reference mining to identify additional relevant papers. This was necessary to capture relevant papers where keywords were not present in the title or abstract.

## 3.2 Delphi Study

A Delphi process comprising two rounds was used to reach expert consensus [65]. The questionnaire for Round 1 is available at https://exaidss.com/wp-content/uploads/2024/04/Questionnaire-Round-1.pdf. The questionnaire for round 2 is available at https://exaidss.com/wp-content/uploads/2024/04/Questionnaire-Round-2.pdf. Participants from the following three groups were invited:

- Group 1  End user decision makers: health professionals, clinicians.
- Group 2  AI Developers: engineers/computer scientists, data scientists, statisticians, implementation scientists, human-computer interaction specialists.
- Group 3  XAI theorists: psychologists/ cognitive scientists, social scientists, philosophers, legal theorists.

The participants were recruited using: (1) an invitation email to experts recommended by the authors; (2) an invitation email to authors of the publications identified through the initial literature search; (3) a call to contribute that was published in a short report (https://exaidss.com/publications/); and (4) the consideration of any expert contacting the lead author on their own initiative. Educational material on XAI and the Delphi study was sent to each participant.

### Objective 1 – An explanation definition for health- AI

The first objective was explored using an iterative process that started with definition fragments identified in the literature and finished using a taxonomical approach, identifying themes in participants' feedback received during the Delphi study. More specifically, in the first round, we divided generic definitions of XAI, identified from the literature, into three components: (1) the semantic entity of an explanation; (2) the aim of the explaining process; and (3) the explanation purpose. Participants were asked to rate on a 1 – 7 Likert scale their agreement on the proposed list of published definitions. Further details can be found at https://exaidss.com/wp-content/uploads/2024/04/Questionnaire-Round-1.pdf Based on the responses from round 1, a more abstract definition was prepared for the second round, asking participants again to rate their agreement using the same scale. More details about the abstract definition can be found at https://exaidss.com/wp-content/uploads/2024/04/Questionnaire-Round-2.pdf. To arrive at the proposed definition of an explanation in health-AI we reviewed participants' feedback obtained during round 2 using a taxonomical approach as applied by McLachlan, *et al.* in [66].

### Objective 2 – Attributes of a good explanation in health-AI

In the second round, the participants were asked to perform the same scoring exercise in a revised item list based on the data gathered from the first round. Only attributes that more than 70% of the participants rated as important (scores 5, 6 and 7) were included in the second round. In addition, participants were presented with two different medical case studies. In both cases participants were provided with two different explanations that varied based on complexity and content. Further details about the explanations provided can be found at https://exaidss.com/wp-content/uploads/2024/04/Questionnaire-Round-2.pdf. Participants were asked to select the explanation they preferred and to justify their choice by identifying reasons from a reasonably extensive pre-defined list of explanation attributes - the same list of attributes that participants reviewed in the previous step. Finally, participants were asked to provide any additional attributes that they would like to see in the provided explanation.

## 4. Results

### 4.1 Scoping Review

Initially, 1052 papers were identified where the prescribed keywords were present in their title or abstract, published after 2018, and in English. After removing duplicates and those papers that did not meet the IC of the study (Table 1), 61 papers were reviewed. Around half of these works included some definition or a list of attributes for a good explanation in health-AI. This was not a formal literature review, but rather a scoping review. The questionnaire for the first round of the Delphi study was developed based on the knowledge drawn from these papers (available at https://exaidss.com/wp-content/uploads/2024/04/Questionnaire-Round-1.pdf).

### 4.2 Delphi Study

From the 135 participants invited to participate in the Delphi study, 39 (29%) participated, of whom: (i) 7 (18%) participated only in the first round; (ii) 14 (36%) participated only in the second round; and (iii) 18 (46%) participated in both rounds. The participants were almost equally distributed between Group 1 – end-user decision makers and Group 2 – AI developers, with fewer participants belonging to Group 3 – XAI theorists (Figure 3). Most had two to six years of experience working on XAI (Figure 4).

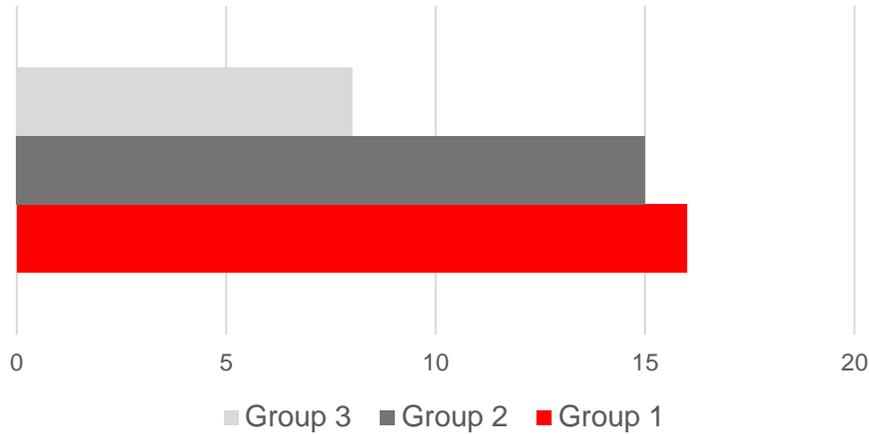

**Figure 2.** Distribution of participants' job profile group

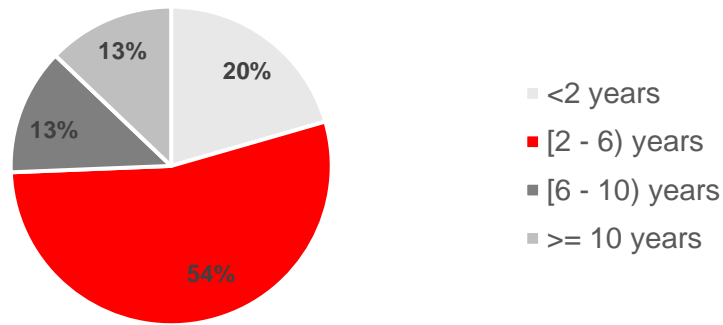

**Figure 3.** Distribution of participants' years of experience working on XA

## Objective 1 – An explanation definition for health- AI

During the first round of the Delphi study, many participants found it difficult to rate their agreement with the proposed definition fragments found in the literature (available at https://exaidss.com/wp-content/uploads/2024/04/Questionnaire-Round-1.pdf). The main criticism was that providing a unique definition for an explanation of health-AI is challenging as it is highly depending on who needs the explanation and for what reasons. Thus, a more abstract definition was proposed in the second round (can be found at https://exaidss.com/wp-content/uploads/2024/04/Questionnaire-Round-2.pdf). As illustrated in Figure 5, 69% of the participants rated their agreement on the proposed abstract definition with rates 5-6. A consistent theme among participants' feedback is that the abstract definition should be less vague and distinguish better between the AI and its output and between the AI and the explanation purpose. Using a taxonomical approach, during which we identified keywords among the participants' feedback, we developed the following definition to answer the first objective ('What is an explanation in health-AI?'):

> *An explanation is a tool intended to assist a user with insights relevant to the function of an AI/ML model, designed for a specific purpose, and the reason for this particular output.*

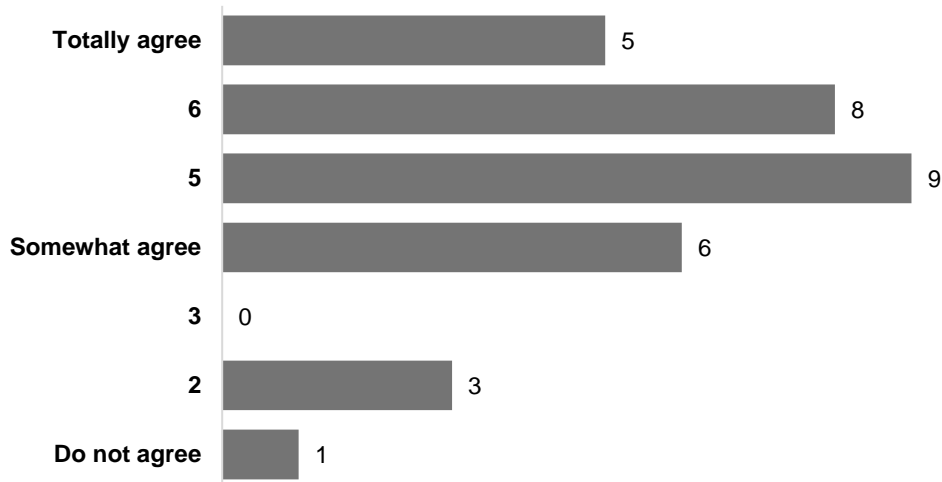

**Figure 4.** Distribution of participants' agreement on the proposed definition of an explanation in health-AI asked during Delphi study round 2.

Each component of our definition is expanded further in Table 2, resulting also in a more concise *alternate definition* consistent with the context and meaning of our definition. Both definitions are read complete from top-to-bottom within their column, with the expanded definition describing the meaning or intent of that element of both definitions.

**Table 2.** Our definition of what an explanation in health-AI is

| Our Definition | Expanded Definition | Alternate Definition |
|---|---|---|
| A tool | An additional component or extension of the AI/ML | A Means |
| Intended to assist | Makes it easier to understand complex AI/ML models | to Support |
| A user | The specific user type operating the AI/ML | an Operator |
| With insights | Information tailored to assist the specific user | with Understandings |
| Relevant to the function of an AI/ML model | How the AI/ML works | for How |
| Designed for a specific purpose | What the AI/ML was designed to do | a focused AI/ML |
| And the reason for | Why the AI/ML produced this particular result | deliberated |
| This particular output | This prediction, probability, recommendation or result | this Result |

## Objective 2 – What are the attributes of a good explanation in health-AI

The attributes of a good explanation were divided into three groups: (1) attributes related to the "focus" of the explanation, (2) attributes related to the "content" of the explanation and (3) attributes related to the "output" of the explanation. Participants rated the importance of each as shown in Figures 6,7 and 8, respectively.

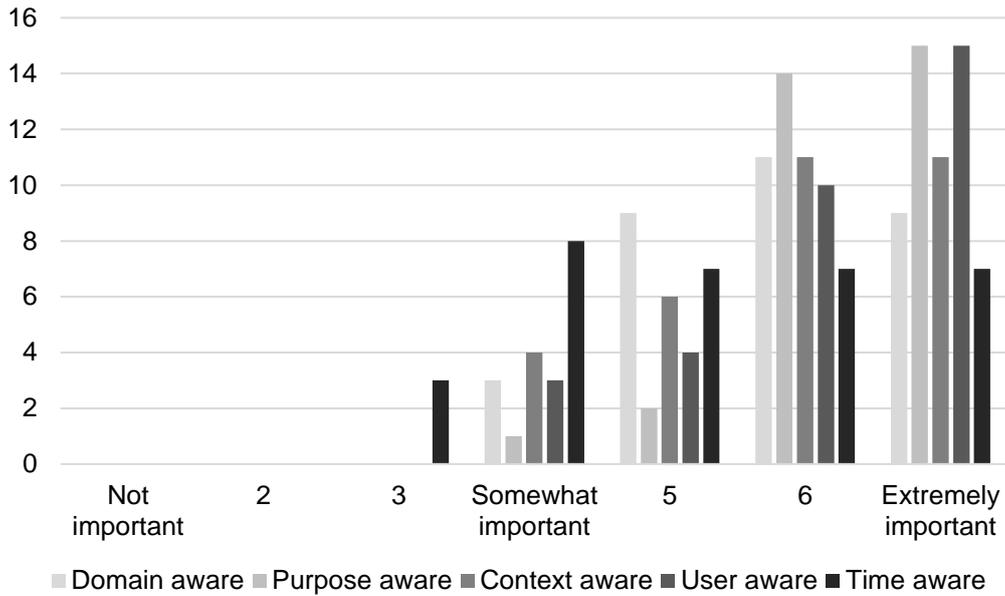

**Figure 5.** Distribution of importance of attributes related to the focus of the explanation of health-AI as proposed in Delphi study round 2.

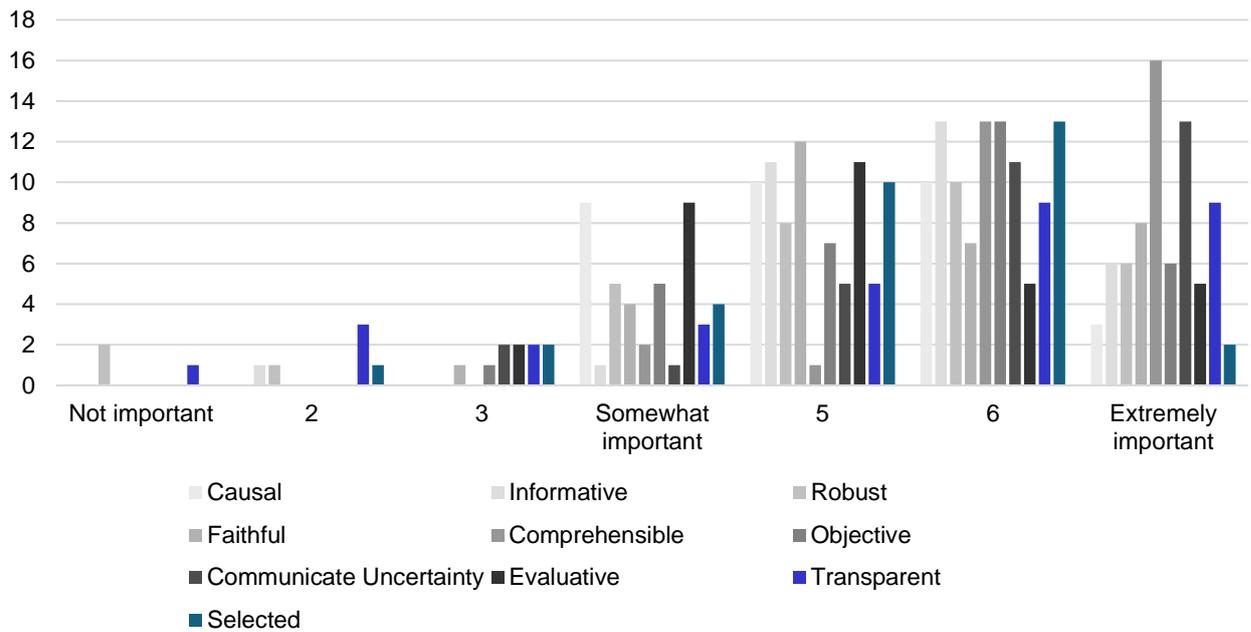

**Figure 6.** Distribution of importance of attributes related to the content of the explanation of health-AI as proposed in Delphi study round 2.

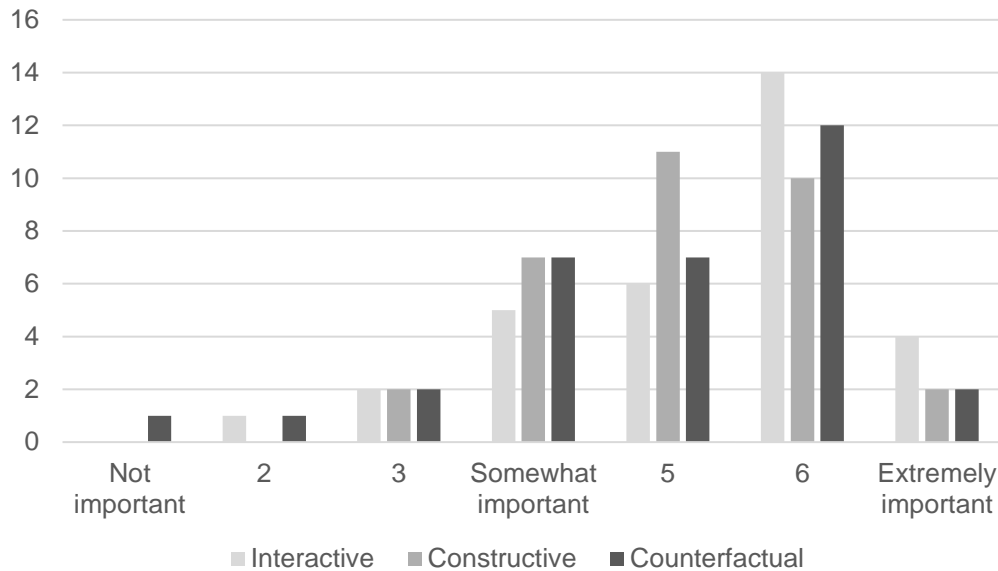

**Figure 7.** Distribution of importance of attributes related to the output of the explanation of health-AI as proposed in Delphi study round 2.

Ratings of 5 to 7 by 80% of the participants or more were defined as "important and critical". Ratings of 5 to 7 by [60% - 80%) of the participants were defined as "desirable but not essential". Ratings of 5 to 7 by less than 60% of the participants were defined as "not important". Based on the participants rates and valuable comments the list of attributes for a good explanation in health-AI, alongside their definitions, which answer the second objective is presented in Table 3. The percentages next to each attribute show how many participants rated each attribute with scores 5 to 7. When provided with two different explanations for the two different case studies, the majority of the participants chose the explanation that was more comprehensible and informative (Explanation B). In Case 1: 88% of participants chose Explanation B and in Case 2: 72% of participants chose Explanation B.

**Table 3.** List of attributes, with their definitions, for a good explanation in health-AI. The percentages next to each attribute show how many participants rated each attribute with scores 5 to 7.

| Attributes | Definition |
|---|---|
| **Important and critical:**<br>1. Purpose aware (97%)<br>2. Comprehensible (94%)<br>3. Informative (94%)<br>4. User aware (91%)<br>5. Domain aware (91%)<br>6. Communicate Uncertainty (91%)<br>7. Context aware (88%)<br>8. Faithful (84%)<br>9. Objective (81%) | 1. Different explanation purposes require different explanations. Therefore, explanations should be tailored to a specific purpose (e.g. an explanation to an AI developer working on improving the AI; an explanation to a patient about their treatment).<br>2. Be clear and understandable to users.<br>3. Provide the necessary and sufficient information to close the user's knowledge gap.<br>4. Be tailored to the user's needs and abilities.<br>5. Be tailored to the domain, incorporating the relevant terms of the domain.<br>6. Explain how certain the prediction is.<br>7. Be relevant to the decision output to be made (e.g. improve patient outcomes, improve use of clinicians time).<br>8. Accurately matches the input-output mapping of the AI system.<br>9. Be as objective as possible to minimize the amount of subjectivity a user might have when interpreting the explanations. |
| **Desirable but not essential:**<br>1. Selected (78%) | 1. Be specific and not consist of the complete cause of an event, highlighting the most important features for a decision. |
| 2. Robust (75%)<br>3. Interactive (75%)<br>4. Causal (72%)<br>5. Transparent (72%)<br>6. Contrastive (72%)<br>7. Evaluative (66%)<br>8. Time aware (66%)<br>9. Counterfactual (66%) | 2. Withstand small perturbations of the input that do not change the output.<br>3. Be a transfer of knowledge, presented as part of a conversation or interaction. Should understand the needs of the user and adapt.<br>4. Provide relevant causal information.<br>5. Help the user in understanding the underlying logic of the AI system, and possibly identifying that the system is wrong.<br>6. Explain questions in the constructive form "Why x and not y?".<br>7. Present evidence to support or refute human judgements and explain trade-offs between any set of options.<br>8. Be tailored to the user's time to engage with the explanation.<br>9. Explain questions in the constructive form "What would happen if?". |

# 5. Discussion

We often measure the performance of AI systems by metrics to determine if it achieves an acceptable performance [12]. However, in healthcare, as in many other fields, high predictive performance is not the only requirement and several unmet needs remain. These unmet needs include lack of explanations in clinically meaningful terms, coping with the unknown medical conditions, and transparency of the system's limitations. All these unmet needs stress the importance of the explainability, transparency, and interactions between the practitioners and the AI systems.

A common theme within the XAI field is to determine when an explanation is proper [38]. Currently, this is hard to achieve as there in no universally agreed definition and list of attributes of a good explanation in AI. A reason for this is that properties of an explanation depend highly on the application domain, the intended purpose of the explanation and user's characteristics. For that reason, we studied expert opinions from a diverse range of interested parties to answer two fundamental questions of explanation in health-AI that remain unanswered: (1) What is an explanation in health-AI? And (2) What are the attributes of a good explanation in health-AI? The research outputs are (1) a definition and (2) a global list of attributes of a good explanation in health-AI.

Regarding objective 1 – *what is an explanation in health-AI?* – we developed our proposed definition using an iterative process that started using definition fragments identified in the literature and finished using a taxonomical approach, identifying themes in participants' feedback received during the Delphi study. Our proposed abstract definition, though not exactly the same, follows the reasoning process described by Saeed et Omlin [17]. They provide an abstract definition of explainability that has three main components: insights, targeted audience and need. Their abstract definition, "explainability aims to help the targeted audience to fulfil a need based on the provided insights from the explainability techniques used", is based solely on a systematic meta-survey without incorporating participants' feedback. The advantage of our proposed definition is that it is based not only on literature, but on participants' feedback from various disciplines. Their feedback was analysed using a well-structured taxonomical approach [66]. The combined keywords identified in participants' feedback resulted in our proposed definition: "*an explanation is a tool intended to assist a user with insights relevant to the function of an AI/ML model designed for a specific purpose and the reason for this particular output*". This definition was developed for health-AI, but it can be applied to other disciplines as well.

Regarding objective 2 – *what are the attributes of a good explanation in health-AI?* – our proposed list of attributes is based on the literature and participants. All the critical and important attributes in our list are identified also in other papers. For instance, the four important attributes regarding the focus of the explanation, "purpose-aware", "user-aware", "domain-aware" and "context aware" have been identified by other researchers using different names such as "purpose oriented" [17], "domain match" [57], "human centeredness", "appropriateness" [55], "audience specific" [15], "user oriented" [46], "nature of user expertise" [33], "context specific" [61], "context sensitive" [67], "context awareness" [14], "contextual" [52]. The five important attributes regarding the content of the explanation, "comprehensible", "informative", "communicate uncertainty", "faithful" and "objective" have been identified by other researchers using different names such as "comprehensibility" [54], [68], [24] "easy to understand" [17], "informativeness" [55], "fidelity" [68], [14], "objectivity", "validity and completeness" [50], "uncertainty awareness" [51], "certainty" [24].

In the remainder of the discussion, we present the lessons learned from participants' feedback, the results implications for theory and practice, the objections against the need for XAI and the limitations, and future research directions.

## 5.1 Lessons learned from experts

***Objective 1 – What is an explanation in health-AI?***

During the first round of the Delphi study, we divided the definition of explainable AI (XAI) into three components (based solely on knowledge identified during the literature review): (1) the semantic entity of an explanation; (2) the aim of the explaining process; and (3) the explanation purpose. For each component, we present published definition fragments and asked participants to rate their agreement. The overwhelming impression from the participants was this part was incomprehensible as it was wrongly assumed that there is single definition. The participants highlighted the fact that the definition of an explanation is influenced by who is using the explanation and for what reason. Therefore, in the second round, we opted for an abstract definition such as "an explanation of AI is an output that assists the user to achieve his/her purpose.", while:

- User: who is intended to use the explanation, such as: doctor, patient, model expert, lawyer, regulator etc.
- Purpose: the purpose for requiring an explanation, such as: increase the trust in model's recommendation, improve the understanding of model's outcome, debug the model, ensure fair and unbiased decisions, inspect the model's properties, etc.
- Output: the type of explanation output, such as: a counterfactual statement, a justification, a list of relevant inputs, etc.

Participants agreed much more with this abstract definition and they provided us with valuable feedback. Some generic comments were provided, such as "while I like the abstract definition, I find it a little vague", while other comments were giving directions on how we can improve our definition, such as:

- "The definition needs to include reference to where the output comes from"
- "It feels like there needs to be some clarification that the explanation is often an additional output or an extension to the output of the AI"
- "The relationship between the explanation and what is being explained"
- "I think the focus should be more on the fact that an explanation should assist the user to understand what and how the AI made the particular prediction/decision and less about what the user might do with that decision"
- "This definition is so general that it seems to make no distinction between the prediction (whether probabilistic or not) and the explanation"

Using a taxonomical approach, we identified common themes among the participants' feedback, which resulted in our proposed definition.

***Objective 2 – What are the attributes of a good explanation in health-AI?***

The participants found the initial list of attributes identified in the literature and presented in the first round of the Delphi quite complete. From the 21 initial attributes of a good explanation, 17 (81%) passed the 70% remark and were included in round 2. In addition, only one new attribute was proposed by the participants in the first round. In the second round, we observed that for all the provided attributes more that 65% of the participants rated them with scores 5 to 7, therefor no attribute was classified as not important. Some useful feedback that we received from the participants during the second round, which could be helpful to other researchers exploring the same fundamental question, were:

- Perhaps a good explanation should be aware of the AI ethical considerations and the AI system's continuous monitoring.
- Attributes such as "interactive," "selected," and "time-aware" were classified as desirable but not essential as they are context—and user-dependent. In other words, they might be considered critical in some circumstances, but not for every context and every user.
- Attributes such as "causal", " contrastive" and "counterfactual" were classified as desirable but not essential because (1) they are context, purpose and user dependant, and (2) participants found that they are not always feasible or easy to achieve, especially when explaining non-causal ML models. The latter was a consistent comment among many AI developers experts, especially for having a causal explanation. This indicates that causality, when it is possible to be achieved, should be considered as critical and not simply desirable.
- Regarding the "evaluative" attribute, one participant commented that a distinction should be made between predictive models that help users understand information (e.g. risk prediction) and models that make judgements (e.g. give advice on decisions). The former type of model is much simpler to explain and leaves the more complex judgements to the user, while the latter is unlikely to be able to capture all scenarios and, therefore, may frequently conflict with a user for valid reasons that are outside of the scope of the model.
- The explanation should be accessible to a diverse audience, considering factors such as language, literacy levels, and any potential sensory limitations.

From the two explanations provided for each case study, participants' feedback indicated that having an informative and comprehensible explanation that communicates uncertainty by indicating the model's confidence in the prediction is very important. These three attributes were also found important during the rating exercise. A frequent feedback from the participants was that they would like the provided explanation to be more interactive, even if this attribute was rated as desirable but not essential, leaving the user the option to ask for more information or to seek an explanation for specific interventional or counterfactual questions. Three more attributes that were not considered in our list and were only mentioned by the participants when selecting the explanation that they preferred the most were that (1) they would prefer a more visual explanation output, (2) they would like the explanation to provide them with possible actions and (3) they would like the explanation to provide details on the training data used and the targeted population.

## 5.2 Implications for theory and practice

We believe that investigating the two fundamental questions that have been neglected: 1- What is an explanation in health-AI? and 2- What are the attributes of a good explanation in health-AI?, will have important implications for different affected parties, such as the:

*XAI research community*: The output of this study will significantly impact the XAI research community by providing consistent fundamental elements for XAI. More specifically, by understanding the attributes of a good explanation, new algorithms for developing improved and more holistic explanations can be developed. However, generating explanations is not enough, it is also crucial to evaluate how good these explanations are. Thus, the proposed list of attributes will also serve as a means to produce an evaluation process, which is currently lacking [69], [70], [71].

*Healthcare professionals and patients*: Having better explanations that are tailored to users' needs will bring XAI closer to adoption in healthcare. Clinicians will benefit directly from this as they will more easily understand how and why the AI provided a specific recommendation. Additionally, it increases the ability of healthcare professionals to better understand the day-to-day patterns and needs of their patients, and with that understanding, they can provide improved personalised care and support for staying healthy. The adoption of AI technologies could potentially be financially beneficial for healthcare providers – such as the UK's NHS – by reducing the cost of consultations, medicines, and resource waste. However, the most important impact will be on patient outcomes, as augmented clinical decision-making delivered with explainable health-AI will improve care.

## 5.3 Objections on the need of XAI and counterarguments

Despite the increased research interest and recent regulation in favour of XAI, many researchers object to the imposition of XAI [72]. Some of the main criticisms against XAI include:

*Loss of accuracy*: One of the main criticisms is that the requirement for explainability might lead to a preference for simpler models, resulting in a loss of accuracy [73]. However, it is also possible that the advocacy for transparency and explainability may lead to a general performance improvement for three reasons: (i) it will help ensure impartiality in decision-making, i.e. to highlight bias in the training dataset; (ii) it will act to improve robustness by highlighting potential adversarial perturbations that could change the prediction; and finally, (iii) it can help ensure that only meaningful variables influence the output, i.e., guaranteeing that an underlying truthful causality exists in the model reasoning [50].

*Loss of AI power*: Expecting humans to review and understand AI reasoning would undermine the key benefits of using AI. AI may yet be far from capable of performing at the level of human cognition and emotional intelligence. However, demanding that the operation of AI 'slow down' so people can follow along in each use case could defeat the entire purpose of using AI [74]. Unfortunately, this perception of XAI is misleading. It is known that AI is powerful as a (self) learning system as it can continuously ingest new information and search for solutions in multiple different and nonlinear ways that may not always be understood by the user. That is why XAI should aim to provide a glimpse into the AI reasoning and not a fully detailed description of how the AI works.

*Neither necessary, nor sufficient to establish trust in AI*: There is a belief that XAI is not necessary for having a trustworthy AI and that a merely accurate AI that is externally validated to show robustness and generalisability is sufficient [75]. However, XAI does not aim to replace model validation (either internal, temporal or external). Both aspects, XAI and validation, are crucial steps needed to establish trust and enable AI usability and integration into the existing workflows [11], [76]. Another belief is that when an AI model produces accurate predictions that aid clinicians to better treat their patients, that model may be useful even without detailed explanations [46]. For example, some AI tools are used to read, interpret and report medical images. However, when AI models are used in an automated fashion, laws and regulations should require an explanation of AI decisions to ensure that they are transparent, fair and capable of reasoned defeasibility [12].

*Not a prerequisite for legitimate and responsible AI*: Finally, few researchers believe that XAI is not a prerequisite for responsible AI. They believe that there are many aspects in our life that we do not fully understand *how* they work, but we accept that they *do* work. They believe AI will be viewed very similarly in the near future [74]. However, we believe this is a simplistic perspective and in high-risk disciplines such as medicine, where patient's lives are affected by AI recommendations, it could even be considered dangerous [76]. While some formative steps are already in place, dynamic regulation similar to that for pharmaceuticals is needed to protect against inaccurate, poorly specific, or intentionally biased AI systems.

### 5.4 Limitations and future research

The first strength of this study is that it aims to answer a fundamental question related to XAI by synthesising published literature and expert opinions. The second strength is that expert opinion was collected using a well-structured Delphi study that comprised two rounds and was supervised by a steering group of suitably qualified academics. Third and finally, the main strength of this study is that the reviewed literature and expert participants in the Delphi study were not limited solely to healthcare, but were selected from three diverse groups: (i) end user decision makers in healthcare; (ii) AI developers; and (iii) XAI theorists. This allowed us to answer the question using an interdisciplinary approach that was lacking in the literature alone, and this work profits from the opinions of these diverse disciplines.

This study has several limitations. First, even if a diverse and sufficient number of participants were included in the Delphi study, having more expert participants would have strengthened our conclusions. It would have also allowed us to study differences between the three types of responders with regard to their rating of attribute importance. While more than 100 participants were invited to participate in this study, only 39 consented and took part in our study. In addition, key decision-makers of regulators, such as administrators/hospital management and policymakers were lacking from our participant cohort. This expert group was initially considered in the study design and 17 regulators were invited to participate, but none responded. Finally, patients were not considered in this study's design but could potentially have offered useful insights regarding what it means for AI to be explainable to their experience as a healthcare consumer. Both regulators and patients should be included in future Delphi studies. Our future research will also seek to go deeper and clarify the unique attributes of a good explanation for different subdomains of medicine (e.g. acute medicine, chronic conditions) and different explanation purposes.

## 6. Conclusion

While the demand for XAI in healthcare is high, determining what constitutes a good explanation is *ad hoc,* and providing adequate explanations remains challenging. Without a solution to the problem of trustworthy AI and user acceptance of healthcare technologies generally, the undeniable benefits of these systems will never be realised, and all our efforts to develop accurate health-AI will be in vain. This research aimed to shed light on two fundamental questions of explanation in health-AI that remain unanswered; (1) What is an explanation in health-AI?, And (2) What are the attributes of a good explanation in health-AI? In this study, we synthesise for the first-time published literature and expert opinions, using a Delphi study, from a diverse research background. The research outputs are (1) a definition and (2) a global list of attributes of a good explanation in health-AI.

**CRediT authorship contribution statement**
**Evangelia Kyrimi:** Conceptualization, Methodology, Formal analysis, Visualization, Writing – Original Draft. **Scott McLachlan:** Conceptualization, Methodology, Resources, Writing - Review & Editing. **Jared M Wohlgemut:** Resources, Writing - Review & Editing. **Zane B Perkins:** Resources, Writing - Review & Editing. **David A. Lagnado:** Conceptualization, Resources, Writing - Review & Editing. **William Marsh:** Resources, Writing - Review & Editing. **ExAIDSS Expert Group:** Resources.

**Competing interests**

The authors declare that they have no known competing financial interests or personal relationships that could have appeared to influence the work reported in this paper.

**Ethics**
This study has been reviewed and approved by QMUL Electronic Engineering and Computer Science Devolved School Research Ethics Committee (QMERC20.565.DSEECS23.052)

**Data availability**
No data was used for the research described in the article.

**Acknowledgement**
E.K is supported by the Royal Academy of Engineering under the Research Fellowship scheme under project ExAIDSS: Explainable AI to ensure trust in clinical Decision Support Systems RF2122-21-226.

# ExAIDSS Expert Group:

**Alexander Gimson[1], Ali Shafti[2], Ari Ercole[3], Amitava Banerjee[4], Ben Glocker[5], Burkhard Schafer[6], Constantine Gatsonis[7], Crina Grosan[12], Danielle Sent[8], David S. Berman[9], David Glass[10], Declan P. O'Regan[11], Dimitrios Letsios[12], Dylan Morrissey[13,14], Erhan Pisirir[15], Francesco Leofante[16], Hamit Soyel[15], Jon Williamson[17], Keri Grieman[18], Kudakwashe Dube[19], Max Mardsen[20,27], Myura Nagendran[21,22,23], Nigel Tai[20,27], Olga Kostopoulou[24], Owain Jones[25], Paul Curzon[15], Rebecca S. Stoner[26], Sankalp Tandle[26], Shalmali Joshi[27], Somayyeh Mossadegh[28], Stefan Buijsman[29], Tim Miller[30], Vince Istvan Madai [31,32]**

---

[1] Cambridge Liver Unit, Addenbrooke's Hospital, Cambridge University Hospitals, Hills Road, Box 210, Cambridge, UK; and New Cambridge Centre for Artificial Intelligence in Medicine (CCAIM), University of Cambridge, UK.
[2] Cambridge Consultants Ltd, UK.
[3] Division of Anaesthesia and Cambridge Centre for AI in Medicine, University of Cambridge, Cambridge, UK; and Magdelene College, University of Cambridge, UK.
[4] Institute of Health Informatics, University College London, London, UK
[5] Department of Computing, Imperial College London, London, UK
[6] SCRIPT Centre for IT and IP Law, School of Law, University of Edinburgh, UK
[7] Center for Statistical Sciences, School of Public Health, Brown University, Providence, USA
[8] Research Group Artificial Intelligence, HU University of Applied Sciences, Utrecht, The Netherlands; and Jheronimus Academy of Data Science, Tilburg University & Eindhoven University of Technology, 's-Hertogenbosch, The Netherlands
[9] Centre for Theoretical Physics, Queen Mary University of London, London, UK
[10] School of Computing, Ulster University, Belfast, UK
[11] MRC Laboratory of Medical Sciences, Imperial College London, London, UK
[12] Department of Informatics, King's College London, UK
[13] Physiotherapy Department, Barts Health NHS Trust, London, UK
[14] Sport and Exercise Medicine, Queen Mary University of London, London, UK
[15] School of Electronic Engineering and Computer Science, Queen Mary University of London, UK
[16] Department of Computing, Imperial College London, UK
[17] Department of Philosophy & Centre for Reasoning, Cornwallis NW, University of Kent, Canterbury, UK
[18] School of Law, Queen Mary University of London, London, UK.
[19] National Language Institute, Midlands State University, Zimbabwe; Institute of Education, Massey University, New Zealand.
[20] Academic Department of Military Surgery and Trauma, Research and Clinical Innovation, Defence Medical Services, Birmingham, UK
[21] UKRI Centre for Doctoral Training in AI for Healthcare, Imperial College London, London, UK
[22] Division of Anaesthetics, Pain Medicine, and Intensive Care, Imperial College London, London, UK
[23] Brain and Behaviour Lab, Imperial College London, London, UK
[24] Department of Surgery and Cancer, Imperial College London, London, UK
[25] Department of Public Health & Primary Care, University of Cambridge, Cambridge, UK
[26] Centre for Trauma Sciences, Blizard Institute, Queen Mary University of London, London, UK
[27] Department of Biomedical Informatics, Columbia University, New York, USA
[28] Hepatic, Pancreatic and Biliary (HPB) Surgery at University Hospital Plymouth NHS Trust, UK; and The Centre for Trauma Sciences, Blizzard Institute, Queen Mary University of London, London, UK.
[29] Delft University of Technology, Delft, The Netherlands
[30] School of Electrical Engineering and Computer Science, The University of Queensland, Brisbane, Australia
[31] QUEST Center for Responsible Research, Berlin Institute of Health (BIH), Charité—Universitätsmedizin Berlin, Germany
[32] Faculty of Computing, Engineering, and the Built Environment, School of Computing and Digital Technology, Birmingham, UK

Errors and omissions excepted